\documentclass[aps,pre,twocolumn,superscriptaddress,showpacs]{revtex4}
\usepackage{epsfig}
\usepackage{amsmath} 
\usepackage{times}

\begin{document}

[Phys. Rev. E {\bf 65}, 065201(R) (2002)]

\title{Cusp-scaling behavior in fractal dimension of chaotic scattering}

\author{Adilson E. Motter}
\affiliation{Department of Mathematics, Center 
for Systems Science and Engineering Research,
Arizona State University, Tempe, Arizona 85287}

\author{Ying-Cheng Lai}
\affiliation{Department of Mathematics, 
Center for Systems Science and Engineering Research,
Arizona State University, Tempe, Arizona 85287}
\affiliation{Departments of Electrical Engineering and Physics,
Arizona State University, Tempe, Arizona 85287}


\begin{abstract}

A topological bifurcation in chaotic scattering is characterized by 
a sudden change in the topology of the infinite set of unstable
periodic orbits embedded in the underlying chaotic invariant set.
We uncover a scaling law for the fractal dimension of the chaotic
set for such a bifurcation. Our analysis and numerical computations
in both two- and three-degrees-of-freedom systems suggest
a striking feature associated with these subtle bifurcations: the dimension
typically exhibits a sharp, cusplike local minimum at the bifurcation.

\end{abstract}
\maketitle

The development of nonlinear dynamics has led to new understanding of important
physical processes, such as chaotic scattering  \cite{Chaos_focus,PG:book}. A
scattering process is chaotic if it exhibits a sensitive dependence on initial
conditions in the sense that, a small change in the input variables before the
scattering can result in a large change in some output  variables after the
scattering. Apparently, in a scattering process, every  physical trajectory is
unbounded, but chaos arises because of a bounded  chaotic set (chaotic saddle)
in the scattering region where, for instance,  the interaction between
particles and potential  occurs, as in a potential scattering system.  
Chaotic saddles are nonattracting and, dynamically, 
they lead to transient chaos \cite{Tel:1990}.
Chaotic scattering is thus the physical manifestation of transient chaos in
open Hamiltonian systems, which occurs
in a variety of  physical contexts \cite{Chaos_focus}. 

An issue of fundamental importance in the study of chaotic scattering is to 
understand how the scattering dynamics changes characteristically
as a system parameter is varied. There are three distinct
classes of bifurcations in chaotic scattering: (1) routes to
chaotic scattering from regular scattering, (2) metamorphic bifurcations 
in which the chaotic saddle suddenly changes its characteristics, and (3)
topological bifurcations in which the topology of the chaotic saddle changes.
Routes to chaotic scattering have been well documented,
including abrupt bifurcations \cite{BGO:1990,Lai:1999}
and saddle-center bifurcation followed by a cascade of period-doubling 
bifurcations \cite{Ding:1990}. There
have also been efforts to study metamorphic bifurcations in chaotic
scattering, such as crisis \cite{LG:1994}. Topological bifurcations,
characterized by topological changes in the subsets of infinite number of
unstable periodic orbits (UPOs) embedded in the chaotic saddles, on the other
hand, have not been well understood. One example of topological  bifurcation
is the massive bifurcation \cite{Ding:1991}.  This type of bifurcations is
{\it subtle} but it is expected to be important in high dimensions, as they
can lead to distinct, physically observable scattering phenomena
\cite{High_d_scattering}. While there have been quantitative scaling results
concerning both routes to chaotic scattering  \cite{BGO:1990,Tel:1991} and
metamorphic bifurcation \cite{LG:1994},  the understanding of topological
bifurcations  in chaotic scattering remains largely at the qualitative level 
\cite{Ding:1991}, particularly in high dimensions \cite{High_d_scattering}.

In this paper, we present a scaling law for the fractal 
dimension of the chaotic saddle in a topological bifurcation that appears
to be typical in both two- \cite{Ding:1991} and three-degrees-of-freedom (DOF)
\cite{High_d_scattering} Hamiltonian systems. We focus on the fractal dimension
because of its importance in shaping physically measurable quantities
such as scattering functions, and because of the fact that most 
analytic scaling results associated with various routes to chaotic scattering 
concern fractal dimension but, to our knowledge, there has been none in the
context of topological bifurcations. Due to the subtlety of a topological
bifurcation,  obtaining quantitative scaling laws, even numerically, is 
highly nontrivial. To be concrete, we state our main result in the context
of potential scattering, with the particle energy $E$ as the bifurcation
parameter. Let $E_c$ be the critical
energy value at which a topological bifurcation occurs in the sense that,
a class of infinite number of UPOs is destroyed and 
replaced by another. The principal result of this paper is then that,
near the topological bifurcation, the fractal dimension of the set of 
singularities in any scattering function (to be described below) scales
with the energy in the following way:
\begin{equation} \label{eq:scaling}
D-D_c \sim | E-E_c|^{D_c},
\end{equation}
where $D_c$ is the dimension value at the bifurcation point. {\it The remarkable
observation is that the dimension-versus-energy curve exhibits a
cusplike minimum at the bifurcation}. We establish Eq. 
(\ref{eq:scaling}) by a scaling argument, and we provide numerical 
verification using a class of 2-DOF scattering systems. 
We also argue that the scaling law should hold in 3-DOF systems.
Our result is interesting, as it represents the first, general, quantitative 
scaling result for topological bifurcation of chaotic scattering.

\begin{figure}
\begin{center}
\epsfig{figure=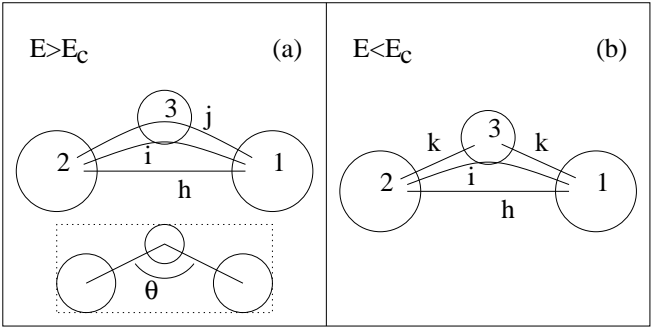,width=\linewidth}
\caption{A topological bifurcation in 2-DOF case: basic
paths composing UPOs (a) before and (b) after the bifurcation.}
\label{fig:2_d_hills}
\end{center}
\end{figure}

To derive the scaling law (\ref{eq:scaling}), we first consider 2-DOF
potential scattering systems. Our idealized, albeit representative,
scattering system consists of three potential hills in the plane, as shown in
Fig. \ref{fig:2_d_hills}. The heights $E_1, E_2$, and $E_3$ of hills
1, 2, and 3 satisfy $E_1, E_2>E_3$. Assume that hill 3  has a circularly
symmetric, quadratic maximum, and that it is close to the line connecting
hills 1 and 2 so that the angle $\theta$ in the triangle formed by the centers
of all three hills is greater than $90^o$, as shown in Fig.
\ref{fig:2_d_hills}(a). When the particle energy $E$ is slightly above $E_3$,
a trapped particle can move back and forth between hills 1 and 2 following
three different paths \cite{Ding:1990}: $h$, $i$, and $j$, as shown in Fig.
\ref{fig:2_d_hills}(a). When the energy drops below $E_3$, the path with
greater deflection by crossing near the center of hill 3, path $j$, is
destroyed because of the appearance of a new forbidden region around the
center of hill 3, and is replaced by two new paths $k$, between hills 1 and 3
and between hills 2 and 3, as shown in Fig. \ref{fig:2_d_hills}(b). For $E
\approx E_3$, the UPOs that form the skeleton of the chaotic saddle are thus
composed of sequences of paths $h, i, j$, and $k$. A topological bifurcation
\cite{Ding:1991} occurs at $E_c = E_3$, at which the class of an infinite
number of  UPOs containing path $j$ is destroyed and replaced by another class
of infinite number of UPOs that contain paths $k$. 

\begin{figure}
\begin{center}
\epsfig{figure=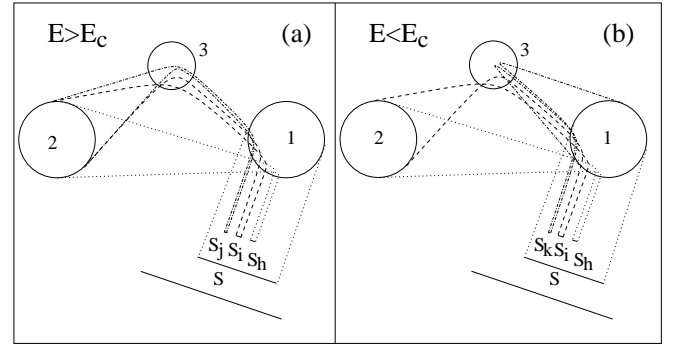,width=\linewidth}
\caption{Construction of the Cantor set.}
\label{fig:cantor_set}
\end{center}
\end{figure}

We focus on a fractal dimension that is physically most relevant: the dimension
$D$ of the set of singularities in a scattering function. Suppose that we
launch particles toward the scatterer from a line outside the scattering
region and measure a dynamical variable characterizing the outgoing particles
after the scattering ($e.g.$, scattering angle), as a function of a variable
characterizing the incoming particles before the scattering ($e.g.$, impact
parameter). Due to chaos, such a scattering function typically contains an
infinite number of singularities that constitute a fractal set, which is the
set of intersecting points between the stable manifold of the chaotic saddle
and the line \cite{Dimension_saddle}. Dynamically, the fractal set of
singularities is constructed by successive interactions of the particles with
the potential hills. For example, consider the interval $s$ of the line
corresponding to particles that are first deflected by hill 1, as shown in Fig.
\ref{fig:cantor_set}. Three subintervals of this interval are then deflected
by the other hills, with gaps between them corresponding to particles that are
scattered to infinity right after the first interaction with hill 1. When
$E>E_c$, these three subintervals are defined by particles that go from hill 1
to hill 2 along paths $h$, $i$, and $j$, respectively, as shown in Fig.
\ref{fig:cantor_set}(a). When $E<E_c$, while two of the subintervals are still
defined by particles that follow paths $h$ and $i$, the third one is now
determined by particles that go from hill 1 to hill 3 and {\it back} to hill 1,
following path $k$, as shown in Fig. \ref{fig:cantor_set}(b). In the next step
each one of the three surviving subintervals will split into three smaller
subintervals, and so on. The splitting due to hill 3 is recorded at successive
interactions with hills 1 and 2.
This process defines a Cantor set with three linear
scales corresponding to the factors by which the intervals are reduced in each
step: $\alpha\approx s_h/s$, $\beta\approx s_i/s$, and $\gamma\approx s_{j}/s$
($E>E_c$) or $\gamma\approx s_{k}/s$ ($E<E_c$), as shown in Fig.
\ref{fig:cantor_set}.

To obtain the dimension $D$ of the Cantor set, we employ the following scaling
argument. Suppose that the critical energy $E_c$, the distance between the
hills, and the effective radii of hills 1 and 2 at $E_c$ are of
order unity. Near the bifurcation, the deflection angle $\phi$ due to hill 3
of the particles corresponding to the edges of the intervals $s_j$
and $s_k$, is smaller and greater than $90^o$ for $E \agt E_c$ and $E \alt E_c$, respectively. In
both cases, $|\tan{\phi}|$ is of order  unity insofar as the angle $\theta$ is
close neither to $90^o$ nor to $180^o$ so that $\tan(\pi -\theta)\sim {\mathcal
O}(1)$ [see Figs. \ref{fig:2_d_hills} and \ref{fig:cantor_set}].
Thus, setting $|\tan{\phi}|$ to be of order unity provides an estimate of
the width of the intervals $s_j$ and $s_k$, and hence an estimate of $\gamma$.
For
concreteness, say the maximum of the potential hill 3 (the {\it bifurcation}
hill) has the following quadratic form: $V = E_c[1 -(r/a)^2]$. A
straightforward calculation \cite{One_hill} indicates that if $|\tan\phi|\sim
{\mathcal O}(1)$, then the impact parameter $b$ scales with
the energy difference: $b\approx b_0 |E-E_c| \equiv b_0\Delta$, where
$b_0$ is proportional to $|\tan\phi|$. Therefore, the Cantor set is determined by the following
scaling factors: $\alpha$, $\beta$, and $\gamma\approx\gamma_0\Delta$, where $\alpha$ and $\beta$
are constants, and $\gamma_0$ is constant on each side of the bifurcation.
Under the approximation that there is a self-similarity in the fractal structure,
the dimension $D$ satisfies the following transcendental equation \cite{BGO:1990}:
$\alpha^D+\beta^D+\gamma_0^D\Delta^D=1.$ For $\Delta$ and $\delta\equiv
D-D(E_c)$ small, to the lowest order, this equation reduces to
\begin{equation} \label{eq:pre_scaling}
[\Delta^{D(E_c)}]^{\delta/D(E_c)}=C\times\frac{\delta/D(E_c)}{\Delta^{D(E_c)}},
\end{equation}
where $C$ is a positive constant. An asymptotic analysis of Eq.
(\ref{eq:pre_scaling})  leads to \cite{Solution} our main scaling result
(\ref{eq:scaling}) \cite{elliptical_hill}.

\begin{figure}
\begin{center}
\epsfig{figure=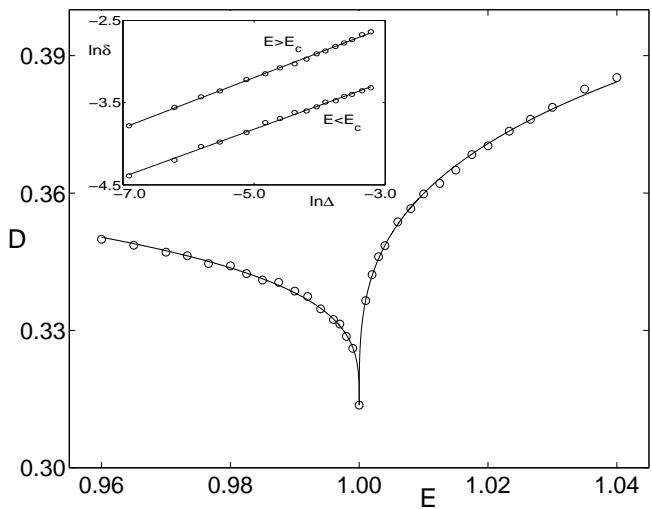,width=\linewidth}
\caption{Numerically obtained cusp scaling behavior in the fractal dimension
about a topological bifurcation.}
\label{fig:dimension_cusp}
\end{center}
\end{figure}

We now present numerical confirmation for the scaling law (\ref{eq:scaling}).
To resolve the cusp structure implied by Eq. (\ref{eq:scaling}) 
is a highly nontrivial task. For numerical feasibility we thus consider 
a planar scattering system consisting of three nonoverlapping
potential hills located at the vertices ($x_i,y_i$) ($i=1,2,3$) of 
an isosceles triangle \cite{Ding:1990}. Each potential is represented by the
following function: $V_i(x,y)= E_i[ 1-(r_i/a_i)^2]$ for $r_i\leq a_i$,
where $r_i=\sqrt{(x-x_i)^2+(y-y_i)^2}$,
and the potential is zero elsewhere. The Hamiltonian is thus:
$H = |{\bf p}|^2/2m + V(x,y)$, where ${\bf p}$ is the two-dimensional 
momentum vector, and $V(x,y) = \sum_{i=1}^{3}V_i(x,y)$. The advantage of utilizing
quadratic potential hills is that the motion of the particle inside each
potential can be solved exactly, making a high-precision dimension
calculation possible. In our numerical experiment, we choose the  following
set of parameter values: $E_1=E_2=10$, $E_3=1$, $a_1=a_2=2$, $a_3=1$,
$(x_1,y_1)=-(x_2,y_2)=(4,0)$, $(x_3,y_3)=(0,2)$, and $m=1$. A topological 
bifurcation thus occurs at $E_c = 1.0$. We utilize the uncertainty algorithm
\cite{MGOY:1985} to compute the dimension $D$ of the set of singularities in
scattering functions. For illustrative purpose, we choose the function to be
the scattering angle to infinity $\psi\equiv
\tan^{-1}{(p_y/p_x)}$ versus the location of the upward (initially) particle on
the horizontal line at $y_0=-4$. The result is shown in Fig.
\ref{fig:dimension_cusp}, in which the cusp behavior is unequivocal
\cite{Computation}. Linear least-squares fits between $\ln{[D(E) - D(E_c)]}$
and $\ln{|E - E_c|}$ yield: $D(E) - D(E_c) = (0.093\pm 0.005)\times
(1-E)^{0.29\pm 0.03}$ for $E<E_c$ and $D(E) - D(E_c) = (0.19\pm 0.01)\times
(E-1)^{0.31\pm 0.03}$  for $E > E_c$, as shown by the inset in Fig.
\ref{fig:dimension_cusp}.   The scaling exponents for both $E > E_c$ and $E <
E_c$ agree well with the theoretically predicted one, the value of 
the dimension at $E_c$, which we obtain numerically: 
$D(E_c) = 0.314 \pm 0.005$. In addition, the proportional factor in Eq.
(\ref{eq:scaling}) is greater for $E > E_c$ than for $E < E_c$,
which is also expected because, theoretically, this factor is 
proportional to $\gamma_0^{D(E_c)}$ and $\gamma_0$ is greater in the former case.
Overall, our computations lend strong
credence to the validity of the scaling (\ref{eq:scaling}).

\begin{figure}
\begin{center}
\epsfig{figure=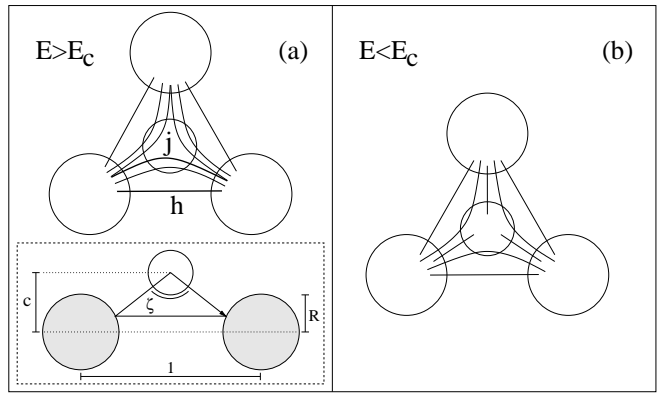,width=\linewidth}
\caption{A topological bifurcation in 3-DOF case: basic
paths (a) before and (b) after the bifurcation.}
\label{fig:3_d_hills}
\end{center}
\end{figure}

Our argument for the scaling law (\ref{eq:scaling}) can in fact be extended
to chaotic scattering in 3-DOF
Hamiltonian systems. As an example, we consider four potential hills
but they are now located in the three-dimensional physical space at the 
vertices of a tetrahedron, as schematically illustrated by planar
projection of the hills in Fig. \ref{fig:3_d_hills}(a) for $E > E_c$
and in Fig. \ref{fig:3_d_hills}(b) for $E < E_c$. We focus on the dimension $D$
of the intersecting set between the stable manifold of the chaotic saddle and a
two-dimensional surface of the phase space, which is the dimension of the set
of singularities in a  scattering function on two variables. Arguments similar
to those in  the 2-DOF case can then be used to derive the
scaling law   (\ref{eq:scaling}) near a topological bifurcation, with the
understanding that the dimension can in principle be either smaller or greater
than 1. When the radii of the potential hills are small, $D(E_c)<1$ can
arise, in which case $dD(E)/dE$ diverges at $E_c$ 
and the dimension exhibits a cusp of the same nature as that in
the 2-DOF case. If $D(E_c)>1$, $dD(E)/dE$ goes   
continuously to zero as $E$ goes to $E_c$, 
which implies that the dimension should be smooth.

Can this smooth behavior be expected for a topological bifurcation in 3-DOF
scattering systems? To obtain an answer, we note the defining characteristic
of a topological bifurcation: the topology of the UPOs is changed but from the
standpoint of symbolic dynamics, no symbol is missing. In our 3-DOF case it
means that, before the bifurcation, all UPOs corresponding to periodic
sequences of the nine basic paths shown in Fig. \ref{fig:3_d_hills}(a) must
actually exist, while after the bifurcation, all those corresponding to
sequences of the paths shown in Fig. \ref{fig:3_d_hills}(b) exist. In
particular, the UPO represented by the sequence $\{h,j\}$ must exist when
$E>E_c$. Without loss of generality, suppose that our 3-DOF system consists of
three hard spheres of radius $R$ at the vertices of a regular triangle with
edge of unit length, and a smooth bifurcation hill located on the symmetrical
axis perpendicular to the center of the triangle. In the inset of Fig.
\ref{fig:3_d_hills}(a) is shown the periodic orbit $\{h,j\}$ for $E>E_c$ in
the limit $E\rightarrow E_c$. A necessary condition for the existence of this
orbit, and hence for the occurrence of the topological bifurcation, is then
$\zeta>90^o$ [see Fig. \ref{fig:3_d_hills}(a): the deflection angle due to
hill 3 is smaller than $90^o$ when $E>E_c$]. In the three-dimensional physical
space, however, there is an additional geometric constraint: $c>1/2\sqrt{3}$,
where $c$ is the orthogonal distance from the center of the bifurcation hill
to the lines connecting the hard spheres [Fig. \ref{fig:3_d_hills}(a)]. We
obtain numerically that the conditions $c>1/2\sqrt{3}$ and $\zeta>90^o$ are
satisfied simultaneously only for $R<0.39$.  Numerical experiments indicate
that, in this range of parameter values,  $D(E_c)$ is smaller than 1. It is
reasonable to assume that the same is valid when the hard spheres are replaced
by smooth potential hills. This implies that for a generic topological
bifurcation in 3-DOF scattering systems, the cusp behavior in Eq.
(\ref{eq:scaling}) is always expected \cite{Counter_example}.

In summary, our scaling analysis and numerical computations reveal a 
striking behavior in the fractal dimension associated with topological
bifurcations in  chaotic scattering: the dimension typically exhibits a cusp
as a function of the bifurcation parameter, with a local minimum at the
bifurcation.

AEM was sponsored by FAPESP.
YCL was supported by AFOSR under Grant No. F49620-98-1-0400 and by NSF
under Grant No. PHY-9996454.


\begin{references}
\bibitem{Dimension_saddle}
The dimension $D$ is related to $d$, the dimension of the chaotic saddle, as:
$d=1+2D$.
\bibitem{One_hill}
The deflection angle $\phi$ due to a potential of the form
$V = E_c[1-(r/a)^2]$ satisfies
$\tan\phi=[b/a]/[\sqrt{(E-E_c)/E_c+(r_0/a)^2}-\sqrt{(r_0/a)^2-(b/a)^2}]$,
where $b$ is the impact parameter and $r_0$ is the initial distance to
the center of the hill. For $|E-E_c|/E_c\ll (r_0 /a)^2$ and $b\ll r_0$
it reads, $\tan\phi\approx [2br_0/a^2]/[(E-E_c)/E_c+(b/a)^2]$.
Therefore, $|\tan\phi|\sim {\mathcal O}(1)$ implies
$b\approx b_0|E-E_c|$, where $b_0= a^2|\tan\phi|/2r_0E_c$.
\bibitem{Solution}
Equation (\ref{eq:pre_scaling}) has the following form: $x^{f(x)} = C f(x)/x$,
where $x = \Delta^{D(E_c)}$ and $f(x) = \delta/D(E_c)$.
For $\lim_{x \rightarrow 0} f(x) \rightarrow 0$, the asymptotic solution
is $x^{f(x)} \rightarrow G$, where $G$ is a constant. Clearly $G$ cannot be 
zero because if it were, then $x$ goes to zero much faster than $f(x)$ does
and, $f(x)/x \rightarrow \infty$, which is inconsistent with the original
equation. In fact, the consistence with Eq. (\ref{eq:pre_scaling}) implies
$G=1$, from which follows $f(x)=x/C$ and hence  Eq. (\ref{eq:scaling}).
\bibitem{Computation}
To obtain Fig. \ref{fig:dimension_cusp}, about 100 h of CPU time is
required on a supercomputer consisting of 17 AMD 1.2 GHz CPUs.
\bibitem{Counter_example}
The scaling relation (\ref{eq:scaling}) is derived for nondegenerate hills.
An instructive counterexample can be constructed as follows. Take the
planar scattering system in Fig. \ref{fig:2_d_hills} and revolve it about the
$y$ axis, which yields a 3-DOF system. The value of the  dimension $D$ is
increased by 1, leading to the following scaling relation: $D-D(E_c)\sim |
E-E_c |^{D(E_c)-1}$. The remarkable  feature is that, despite $D(E_c)>1$,
the cusp at $E_c$ is still present.
\end{references}
\end{document}